\documentclass[prl,twocolumn,showpacs]{revtex4}  
\usepackage{graphicx}  

\begin{document}


\title{Volume Reflection and Volume Refraction of Relativistic 
Particles in a Uniformly Bent Crystal \footnote{ JETP Letters, 2008, Vol. 87, No.2, pp. 87 -91} }

\author{G.\,V.\,Kovalev \/\thanks}



\affiliation{North Saint Paul,  MN 55109, USA }


\date{Dec. 3, 2007}

\begin{abstract}
The scattering of fast charged particles in a bent crystal has been analyzed in the framework of relativistic classical mechanics. The expressions obtained for the deflection function are in satisfactory agreement with the experimental data for the volume reflection of relativistic protons obtained in [1,2,3]. The features of the scattering of the particles on ring potentials are considered in a wide range of impact parameters.
\end{abstract}

\maketitle

In the 1980s, studying the effect of the volume cap-
ture of relativistic protons into the channeling regime,
Taratin and Vorobiev [4, 5] demonstrated the possibility
of volume reflection, i.e., the coherent small-angle scattering of particles at angle $\theta \; \tilde{<} \; 2\theta_L$ ($\theta_L$
 is the Lindhard critical angle) to the side opposite to the bending of the
crystal. Recent experiments reported in [1–3] confirm
the presence of this effect for 1-, 70-, and 400-GeV proton beams in a Si crystal. The conclusions made in [4,
5] were based primarily on the numerical simulation. In
view of this circumstance, the aim of this work is to
derive analytical expressions for the deflection function
of relativistic particles. At first sight, the perturbation
theory in the potential can be applied at relativistic
energies and weak crystal potential [$U(r)\approx 10-100$]. However, the relativistic generalization of the known classical formula for small-angle scattering in
the central field [6],
\[
\chi = -b\int^{\infty}_{b}\frac{d\phi}{d r} \frac{ d r}{\sqrt{r^2-b^2}}, 
\]
where $b$  is the impact parameter and  $\phi(r)=\frac{2U(r) E}{p_{\infty}^2 c^2}$, $U(r), E, p_{\infty}$  are the centrally
symmetric continuous potential of bent planes, total
energy, and particle momentum at infinity, respectively,
is inapplicable for the entire range of impact parameters. Indeed, the above formula is the first nonzero term of the expansion of the classical deflection function
\begin{eqnarray}
\chi(b)=\pi-2b\int^{\infty}_{r_{o}}\frac{d r }{r \sqrt{r^2[1-\phi(r)]-b^2}},
\label{deflection_function}
\end{eqnarray}
in the power series in the “effective” interaction potential  $\phi(r)$. The crystal interaction potential  $U(r)$ is the
sum of the potentials of individual bent planes concentrically located in the radial direction with period $d$.  It has no singularities (i.e., is bounded in magnitude) and
is localized in a narrow ring region at distances $R-N d <r< R$
 (where the crystal thickness $N d << R$ and $N$ 
 is the number of planes). In this region, $U(r)>0$ and $U(r)< 0$ for the positively and negatively charged scattered particles, respectively. Beyond the ring region, it
vanishes rapidly. The perturbation theory in the interaction potential is obviously well applicable if the impact
parameter satisfies the inequality $b < (R-Nd)$.  In this
case, the scattering area localized in the potential range
is far from the turning point  $r_o$  determined from the
relation  $b=r_o \sqrt{1-\phi(r_o)}$  and the root singularity of the
turning point does not contribute to integral (1). In the
general case, it can be shown [7, 8] that the condition of
the convergence of the power series of $\phi$
 is a monotonic 
increase in the function $u(r) = r\sqrt{1-\phi(r)}$ (e.i.  $u(r)^{'} > 0$) in the $r$ 
 regions substantial for integral (1). Such a
monotonicity is achieved if the energy and momentum
of the relativistic particle satisfy the inequality
\begin{eqnarray}
\frac{p_{\infty}^2c^2}{2E}> U(r)+\frac{r}{2} U(r)^{'}.
\label{OrbittingCondition}
\end{eqnarray}
The derivation of this condition is omitted, because it
was given in the Appendix in [8], but the nonrelativistic
case was considered in that work. Taking into account
that the inequality  $U(r)<< r U(r)^{'}$ is satisfied for large
distances  $r\approx R$, relation (2) is transformed to the known
Tsyganov criterion [9] [$R < \frac{p_{\infty}^2c^2}{E U(r)^{'}}$] of the disappearance of channeling in a strongly bent crystal. Thus,  the perturbation theory is obviously applicable only for
strongly bent crystals, where channeling is absent. It is
interesting that the nonrelativistic variant [8] of condition (2) corresponds to the criterion of the absence of
the so-called spiral scattering appearing in small-energy chemical reactions [10].
For this reason, the exact solution of the problem of
the relativistic scattering on the model potential of the
periodic system of rectangular rings
\begin{eqnarray}
U(r)= U_{o} \left\{
\begin{array}{ll}
	1,&  R-i d-a < r < R-i d,  \\
	0,&  R-(i+1)d < r < R-i d-a,\\ 
	0,&	 r < R-Nd, r > R,
\end{array} \right.
\label{Potential_M}
\end{eqnarray}
where $i=0,1,...,(N-1)$ and $a$ and $d$ is the thickness
of a single plane and the interplanar distance, is considered below. With the use of the representation
\begin{eqnarray}
\frac{\pi}{2}=b\int^{\infty}_{r_{o}}\frac{d r }{r \sqrt{r^2[1-\phi(r_{o})]-b^2}},
\end{eqnarray}
particle deflection function (1) can be rewritten in the
form
\begin{eqnarray}
2b\int^{\infty}_{r_{o}}\frac{d r }{r}(\frac{1}{ \sqrt{r^2[1-\phi(r_o)]-b^2}}-\frac{1}{ \sqrt{r^2[1-\phi(r)]-b^2}}).
\label{deflection_functionM1}
\end{eqnarray}
Let us consider the range of small impact parameters,
 $b < R-Nd$, when particles penetrate into the inner
region of the potential ring and twice intersect each
plane. This geometry is not used in the experiments,
because it requires very thin crystals. However, such an
approach to solve the problem makes it possible to easily calculate the exact deflection function on potential
(3) for all impact parameters and, moreover, to demonstrate an interesting property of the scattering on potentials with an empty core. This property does not seemingly attract particular attention. In this case, the turning point is located in the inner region of the potential
ring, where $\phi(r_o)=0$, and, as easily seen from the equation  $b=r_o \sqrt{1-\phi(r_o)}$
, the shortest distance from the
center is $r_o=b$ (see Figs. 1C and 1D).

\begin{figure}
	\centering
		\includegraphics{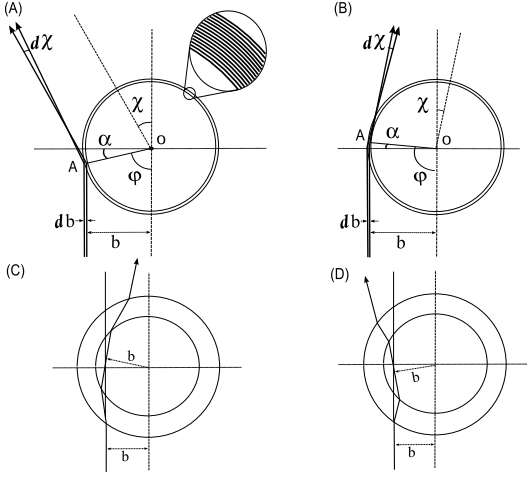}
	\caption{ Scattering of the charged particles on the ring potential: (A) reflection,  $\chi>0$, and (B) refraction,  $\chi<0$. The
effect of the empty core for the (C) positively and (D) negatively charged relativistic particles.}
	\label{fig:ScatteringRing}
\end{figure}

The deflection function for such impact parameters
has the form
\begin{eqnarray}
\chi(b)=2b\int^{\infty}_{b}\frac{d r }{r}(\frac{1}{ \sqrt{r^2-b^2}}-\frac{1}{ \sqrt{r^2[1-\phi(r)]-b^2}}).
\label{deflection_functionM2}
\end{eqnarray}  
Let us consider an arbitrary positive centrally symmetric potential $\phi(r)\geq 0$. Then, the inequality $\frac{1}{ \sqrt{r^2-b^2}} \leq \frac{1}{ \sqrt{r^2[1-\phi(r)]-b^2}}$
 is always valid in the integrand in Eq. (6). As immediately follows from this inequality,
the scattering angles for impact parameters  $0 < b\; \tilde{<} \; (R-Nd)$ are always negative, $\chi < 0$ for any form of the
positive potential $\phi(r)$;  i.e., an arbitrary positive potential is attractive. On the contrary, an arbitrary negative potential, $\phi(r)\leq 0$, is repulsive. This seeming paradox
is called the empty core effect. It is schematically
shown in Figs. 1C and 1D and in Figs. 2A–2D for the scattering on the rectangular potential and can be
treated in various ways, one of which is as follows. The
integral action of the positive potential deflects a particle passing through it to the left from the initial direction (see Fig. 1A); i.e., the particle is reflected. The
potential in the core is absent and, according to the conservation laws, the absolute value of the momentum
and angular momentum should coincide with the
respective initial values. Therefore, the particle trajectory in the core should touch a circle with the radius
equal to the impact parameter. This is possible only if
the particle intersecting the inner boundary of the
potential is deviated to the right from the initial direction. The total deflection angle is equal to the doubled
angle to the turning point. Therefore, the total rotation
is clockwise; i.e., the particle is attracted to the center.
The opposite situation occurs for the negative potential.

Let us introduce the convenient notation
\begin{eqnarray}
\Phi = 1-\phi_o, \; \; \phi_o=\frac{2U_{o} E}{p_{\infty}^2 c^2},\; \;
\hat{r}=\frac{r}{R}, \; \;\hat{b}=\frac{b}{R}, \; \; \hat{a}=\frac{a}{R},  \nonumber\\
\hat{d}=\frac{d}{R}, \; \; \hat{b}_{i} =\frac{\hat{b}}{1-i \hat{d}},\; \; \hat{b}_{a} =\frac{\hat{b}}{1-\hat{a}}, \; \; \hat{b}_{ia}=\frac{\hat{b}}{1-\hat{a}-i \hat{d}}.
\label{definitions}
\end{eqnarray}  
For potential (3), the scattering problem is solved
exactly and deflection function (6) for b < R – Nd is represented in the form of the sum,
\begin{eqnarray}
\chi(b)=2 \alpha(b)=2 \sum_{i=0}^{N-1} \alpha_i(b),
\label{Deflection_General_Crystal} 
\end{eqnarray}
of the integrals over the regions filled with the potential:
\begin{eqnarray}
\alpha_i(b)=\hat{b}\int^{1-i\hat{d}}_{1-i\hat{d}-\hat{a}}\frac{d \hat{r} }{\hat{r}}(\frac{1}{ \sqrt{\hat{r}^2-\hat{b}^2}}-\frac{1}{ \sqrt{\Phi \hat{r}^2-\hat{b}^2}}).
\label{Deflection_i_Ring}
\end{eqnarray}  
Integral (9) is easily calculated and, which is most
important, has an analytic continuation valid for any
impact parameter:
\begin{eqnarray}	
\alpha_{i}(b) =arcsin(\frac{\hat{b}_{i}(\sqrt{1-\hat{b}^2_{i}}-\sqrt{\Phi-\hat{b}^2_{i}})}{{\sqrt{\Phi}}})-\nonumber\\ arcsin(\frac{\hat{b}_{ia}(\sqrt{1-\hat{b}_{ia}^2}-\sqrt{\Phi-\hat{b}_{ia}^2})}{{\sqrt{\Phi}}}).
\label{DeflectionGen_i}
\end{eqnarray}
Note that only the real parts of the deflection function
are meaningful. Therefore, if the impact parameter $\hat{b}$ is
such that any root in Eq. (10) is imaginary, it should be
rejected. In what follows, an approximation that is not
strictly necessary is used. All below formulas can be
derived without this approximation, but it significantly
simplifies the form and use of all of the expressions, is
accepted. For small angles entering into Eq. (10) and
$\Phi \approx 1$ in the denominators in Eq. (10), Eq. (10) can be
represented in the form
\begin{eqnarray}	
\alpha_{i}(b) =\hat{b}_{i}(\sqrt{1-\hat{b}^2_{i}}-\sqrt{\Phi-\hat{b}^2_{i}})-\nonumber\\ \hat{b}_{ia}(\sqrt{1-\hat{b}_{ia}^2}-\sqrt{\Phi-\hat{b}_{ia}^2}).
\label{DeflectionGen_i_app}
\end{eqnarray}
Let us consider the deflection function given by
Eq. (10) for one ring ($N = 1$ and $i = 0$) in more detail:
\begin{eqnarray}	
\alpha(b) =arcsin(\frac{\hat{b}(\sqrt{1-\hat{b}^2}-\sqrt{\Phi-\hat{b}^2})}{{\sqrt{\Phi}}})-\nonumber\\ arcsin(\frac{\hat{b}_{a}(\sqrt{1-\hat{b}_{a}^2}-\sqrt{\Phi-\hat{b}_{a}^2})}{{\sqrt{\Phi}}}).
\label{DeflectionFunction_Ring_One}
\end{eqnarray}
If the inner ring radius is equal to zero, then $\hat{a}=1$ and
last two radicals in Eq. (12) become minimal. According to the above rule, they should be rejected. The
resulting exact function describes the scattering of the
particles on a rectangular cylindrical disc ($U_o >0 $) or
well ($U_o <0 $) [6] (see Fig. 2).

\begin{figure}
	\centering
		\includegraphics{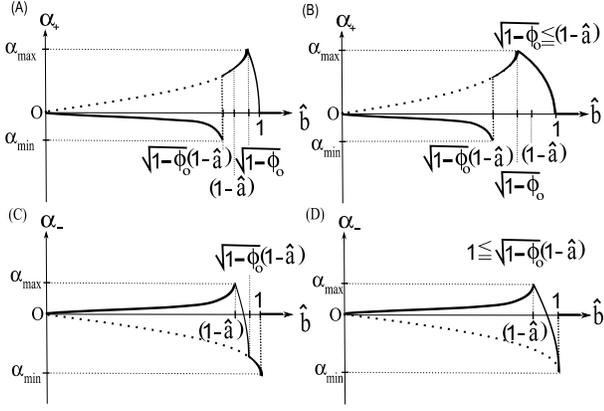}
	\caption{Deflection function $\alpha(\hat{b})$ for positively charged particles from the relativistic particle at (A)  $\hat{a} > \phi_o/2$ and  (B)
  $\hat{a} < \phi_o/2$ and for negatively charged particles at (C) 
 $\hat{a} > |\phi_o|/2$, (D)- $\hat{a} < |\phi_o|/2$. The dotted lines correspond to the
scattering on (A and B) a disc ($U_o > 0$) and (C and D) a well
($U_o < 0$). The region of the empty core effect for positively
charged particles $0 < \hat{b} <\sqrt{\Phi}(1-\hat{a})$ is smaller than the
inner radius $(1-\hat{a})$.}
	\label{fig:DeflectionFunctionF}
\end{figure}

Furthermore, let us analyze the range of impact parameters in the ring region
$(1-N\hat{d})\leq \hat{b}\leq 1$, where    $\hat{b}_{i}$ and $\hat{b}_{ia} \cong 1$ 
 and can be taken
due to the smallness   $N\hat{d}<<1$ and these factors in
Eq. (11) can also be omitted. In this approximation,
Eq. (12) has the form
\begin{eqnarray}	
\alpha(b) =(\sqrt{1-\hat{b}^2}-\sqrt{\Phi-\hat{b}^2})-\nonumber\\ (\sqrt{1-\hat{b}_{a}^2}-\sqrt{\Phi-\hat{b}_{a}^2}).
\label{DeflectionFunction_Ring_app}
\end{eqnarray}

For various $\phi_o$ values ($\phi_o$ is the square of the Lindhard
critical angle with the sign of the charge of the scattered
particle), there are two sequences of critical points that
can pass through the impact parameter  $\hat{b}$  when increasing from 0 to 1. For the positive potential $U_o > 0$ ($\Phi < 1$), the critical points form a sequence $\sqrt{\Phi} (1-\hat{a})<  \sqrt{\Phi} < 1$. Thus, the deflection function for positively charged particles on one ring has the form
\begin{equation}
\alpha(\hat{b})_{+}=
\label{DeflectionFunction_Ring_v1}
\end{equation}
\begin{eqnarray}
\left \{
\begin{array}{ll}
(\sqrt{1-\hat{b}^2}-\sqrt{\Phi-\hat{b}^2})-(\sqrt{1-\hat{b}_{a}^2}-\sqrt{\Phi-\hat{b}_{a}^2}),& \\  for \; \; 0 < \hat{b} <\sqrt{\Phi}(1-\hat{a});\nonumber\\ 
\sqrt{1-\hat{b}^2}-\sqrt{\Phi-\hat{b}^2}, \; \;for \; \;\sqrt{\Phi}(1-\hat{a}) < \hat{b}< \sqrt{\Phi};\\ 
\sqrt{1-\hat{b}^2},\; \;for\; \;  \sqrt{\Phi} < \hat{b}< 1;\\
0, \; \;for \; \;1 < \hat{b}.
\end{array} \right.
\end{eqnarray}
For negative potential $U_o < 0$ ($\Phi > 1$), the critical points
form another sequence $(1-\hat{a})<(1-\hat{a})\sqrt{\Phi}<1$. It provides the deflection function for the scattering of negatively charged particles on one ring:
\begin{equation}
\alpha(\hat{b})_{-}=
\label{DeflectionFunction_Ring_v2}
\end{equation}
\begin{eqnarray}
\left \{
\begin{array}{ll}
(\sqrt{1-\hat{b}^2}-\sqrt{\Phi-\hat{b}^2})-(\sqrt{1-\hat{b}_{a}^2}-\sqrt{\Phi-\hat{b}_{a}^2}),& \\  for \; \; 0 < \hat{b} <(1-\hat{a});\\  
\sqrt{1-\hat{b}^2}-\sqrt{\Phi-\hat{b}^2}+\sqrt{\Phi-\hat{b}_{a}^2},\nonumber\\for \; \;(1-\hat{a}) < \hat{b} <(1-\hat{a})\sqrt{\Phi};\nonumber\\ 
\sqrt{1-\hat{b}^2}-\sqrt{\Phi-\hat{b}^2},\; \;for\; \; (1-\hat{a})\sqrt{\Phi} < \hat{b}< 1;\\
0, \; \;for \; \;1 < \hat{b}.
\end{array} \right.
\end{eqnarray}
Figures 2A and 2C show deflection functions (14) and
(15), respectively, for $|\phi_o|/2 < \hat{a}$ . If the potential is sufficiently large, $|\phi_o|/2 > \hat{a}$ , the arc of the reflection of the
positively charged particles from the outer wall of a
bent plane extends to the left and can be larger than the
width of the ring (and the distance between rings if the
system of rings is considered). This effect causes the
reflection of relativistic particles in the crystal. For negatively charged particles, when $|\phi_o|/2 > \hat{a}$, the corresponding reflection from the inner wall of the potential
is shifted to the right and, correspondingly, the impact
parameter region for the refraction of negatively
charged particles is narrowed. However, at least the narrow refraction region always exists. These two variants
are shown in Figs. 2B and 2D. The substitution of  $\hat{b}=\sqrt{\Phi}$,  $\hat{b}=\sqrt{\Phi}(1-\hat{a})$   and $\hat{b}=(1-\hat{a})$, $\hat{b}=1$  into
Eqs. (14) and (15), respectively, provides the maximum
(minimum) deflection angles for the positively and negatively charged particles, respectively:
\begin{eqnarray}	
\alpha_{max +} =-\alpha_{min -}=\sqrt{|\phi_o|},& &\nonumber\\ 
\alpha_{min +} =-\alpha_{max -} =\frac{|\phi_o|}{2\sqrt{2\hat{a}}} -\sqrt{|\phi_o|}.
\label{MaxMin}
\end{eqnarray}
The deflection functions for the system of bent planes
forming the crystal are similarly obtained from Eqs. (8)
and (11). In this case, the summation in Eq. (8) should
be performed from $0$ to $k – 1$, where $k$ is the ordinal
number of the radial period containing the turning
point. The sequences of the critical points mentioned
before Eqs. (14) and (15) and reflection functions given
by Eqs. (14) and (15) refer to the $k$-th period. The formulas for the reflection functions are omitted in this short
paper, but the corresponding plots are presented in
Fig. 3.
Two length parameters,  and $\hat{a}$, $\hat{d}$ ($\hat{a}<\hat{d}$), exist in
the periodic system of bent planes. Hence, three different variants of the curves can exist with (i)   $\hat{d} > \hat{a} > \phi_o/2$, (ii)  $\hat{d}> \phi_o/2 > \hat{a}$, and (iii) $\phi_o/2 >\hat{d}>  \hat{a}$. Figure 3
shows the final result for the reflection function in the first and third variants.

\begin{figure}
	\centering
		\includegraphics{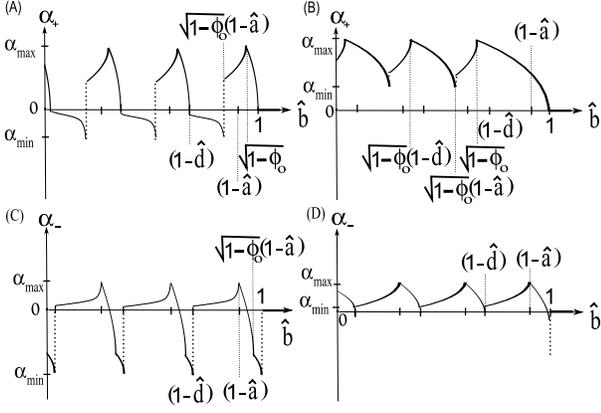}
	\caption{Deflection function for the ring crystal and positively
charged particles at (A)  $\hat{a} > \phi_o/2$ and (B) $\hat{d} < \phi_o/2$ and for
negatively charged particles at (C)  $\hat{a} > |\phi_o|/2$ and $\hat{d} < |\phi_o|/2$.}
	\label{fig:DeflectionFunctionS}
\end{figure}

As mentioned above, under the condition
\begin{eqnarray}	
\phi_o >\frac{2d}{R},
\label{ConditionOfReflection}
\end{eqnarray}
the refraction regions of positively charged particles
disappear in the pattern of the deflection function (see
Fig. 3B). Since this effect is of an applied interest for
controlling the relativistic particle beams, let us calculate the average reflection angle under these conditions.
For a rough estimate, the extreme ring where the deflection function has the simplest form of Eq. (14) is used.
The average deflection angle in the impact parameter
range $\sqrt{\Phi} < \hat{b}< 1$  is determined by the integral 
\begin{eqnarray}	
\bar{\alpha}_{+} = \frac{1}{1-\sqrt{\Phi}} \int^{1}_{\sqrt{\Phi}} 
\sqrt{1-\hat{b}^2} d\hat{b} \cong \frac{2\sqrt{\phi_o}}{3}.
\label{AverageAnglePReflectLast}
\end{eqnarray}
Thus, the average reflection angle is $\chi_{+}=2\bar{\alpha}_{+}=4\sqrt{\phi_o} / 3$, which is equal to $1.33\cdot \theta_L$. A more accurate estimate of the reflection angle can be obtained by averaging over any inner period of the reflection function. Let
us use the second impact parameter period from the
edge, $(1-\hat{d})\sqrt{\Phi} < \hat{b}< \sqrt{\Phi}$
 (see Fig. 3B).   In this case,
it is unnecessary to calculate the total sum in Eq. (8); it
is sufficient to calculate the reflection function at two
extreme rings. This problem is solved by calculating
two integrals
\begin{eqnarray}	
I_{1} = \int^{(1-\hat{a})\sqrt{\Phi}}_{(1-\hat{d})\sqrt{\Phi}} 
\Bigl((\sqrt{1-\hat{b}^2}-\sqrt{\Phi-\hat{b}^2})-\nonumber\\
(\sqrt{1-\hat{b}^2_{a}}-\sqrt{\Phi-\hat{b}^2_{a}})+\nonumber\\ \sqrt{1-\hat{b}_{1}^2}\Bigl) d\hat{b},
\label{AverageAnglePI1}
\end{eqnarray}
and
\begin{eqnarray}	
I_{2} = \int^{\sqrt{\Phi}}_{(1-\hat{a})\sqrt{\Phi}} 
\Bigl(\sqrt{1-\hat{b}^2}-\sqrt{\Phi-\hat{b}^2}\Bigl) d\hat{b},
\label{AverageAnglePI2}
\end{eqnarray}
with subsequent expansion in small parameters $\hat{a},\hat{d}$ and  $\phi_o$. As a result, the average deflection angle 
$\bar{\alpha}_{+}=(I_1+I_2)/(\hat{d}\sqrt{\Phi})$ is obtained in the form
\begin{eqnarray}	
\bar{\alpha}_{+} = \frac{1}{3\hat{d}}\Bigl(\phi_o^{3/2}+ (2\hat{d}+\phi_o)^{3/2}+(2\hat{d}-2\hat{a})^{3/2}-\nonumber\\-2\sqrt{2}\hat{d}^{3/2}-(2\hat{d}-2\hat{a}+\phi_o)^{3/2}-(2\hat{a}-2\hat{d}+\phi_o)^{3/2}\Bigl).
\label{AverageAnglePReflectF}
\end{eqnarray}

The experiment with 1-GeV protons in a $<\!111\!>$ Si crystal with a bending radius of $R=0.33$m provides an average reflection angle of $236 \mp 6.0$ $\mu$rad [2]. In view
of the data  $\phi_o=\theta_{L}^{2}=0.289\cdot10^{-7}$, $a=0.78 \AA$, $d=3.136 \AA$, Eq. (21) provides $\chi_{+}=2\cdot\bar{\alpha}_{+}=318.8$  $\mu$rad.   

Note that rough formula (18) provides the value $\chi_{+}=2\cdot\bar{\alpha}_{+}=226.6$ $\mu$rad  that is closer to the measured value. The
experiment with 70-GeV protons in a  $<\!111\!>$ Si crystal
with a bending radius of $R = 1.7$ m provides an average
reflection angle of $39.5\mp2.0$ $\mu$rad.   
In view of the data
$\phi_o=\theta_{L}^{2}=0.58\cdot10^{-9}$, $a=0.78 \AA$, $d=3.136 \AA$,
Eq. (21) provides$\chi_{+}=37.3$ $\mu$rad, which is close to the
experimental value. Formula (18) provides a smaller
angle of 32.0 $\mu$rad.

For the latest CERN experiment for
the reflection of 400-GeV protons [3] in a Si crystal oriented in the  $<\!110\!>$ direction ($R = 18.5$ m,  $\phi_o=\theta_{L}^{2}=0.1132\cdot10^{-9}$, $a=0.48 \AA$, $d=1.92 \AA$) and the  $<\!111\!>$ direction ($R = 11.5 $m, $\phi_o=\theta_{L}^{2}=0.1008\cdot10^{-9}$, $a=0.78 \AA$, $d=3.136 \AA$), Eq. (21) provides $\chi_{+}=2\cdot\bar{\alpha}_{+}=19.0$
and $16.0$  $\mu$rad, respectively.  Rough estimate (18) provides values $14.1$ and $13.3$  $\mu$rad, which is very close to
the experimental values $13.9\mp0.2$ and $13.0$  $\mu$rad, for
the  $<\!110\!>$ and  $<\!111\!>$ orientations, respectively.

Comparison of the theoretical estimates and experimental data shows that estimate formula (18) yields a smaller reflection angle over the entire range of
scanned energies than that assumingly more accurate
formula (21). The experimental data reported in [1, 11]
are sufficiently well reproduced by formula (21), but
the results reported in [2, 3] are closer to estimate (18).
This circumstance requires a more detailed analysis of
the experimental conditions and the accuracy of measurements.

I am grateful to Yu.M. Ivanov for the possibility of
a reading preprint of [11].


\end{document}